\pdfoutput=1
\documentclass{PoS}
\usepackage[]{units}
\usepackage{subcaption}
\usepackage{amsmath}

\title{Astrophysical interpretation of small-scale neutrino angular correlation searches with IceCube}

\ShortTitle{Neutrino angular correlations in IceCube}

\author{\speaker{Michael Schimp}, Martin Leuermann and Christopher Wiebusch\\
        III. Physikalisches Institut, RWTH Aachen University, 52056 Aachen, Germany\\
        E-mail: \email{michael.schimp@rwth-aachen.de}}

\abstract{IceCube, a cubic-kilometer sized neutrino detector at the Geographic South Pole, has recently discovered a diffuse all-flavor flux of astrophysical neutrinos.
However, the corresponding astrophysical sources have not yet been identified in current IceCube analyses.
We present a method to interpret the results of a recently published angular correlation analysis in IceCube searching for spatial clustering of muon neutrino events in terms of astrophysical models (given by an arbitrary source count distribution).
We exemplarily show the resulting limits on the parameters of a class of source count distributions motivated by Fermi-LAT observations of resolved blazars.}

\FullConference{The 34th International Cosmic Ray Conference,\\
		30 July- 6 August, 2015\\
		The Hague, The Netherlands}

\begin{document}
\section{Introduction}
IceCube is a cubic-kilometer neutrino detector at the Geographic South Pole capable to detect neutrinos in the energy regime above $\sim \unit[10]{GeV}$.
It consists of 86 strings in the Antarctic glacial ice, each equipped with 60 photosensors deployed at depths between $\unit[1450]{m}$ and $\unit[2450]{m}$ below the surface \cite{bib:IC}.
The detection of high-energy neutrino events is based on the measurement of Cherenkov photons, emitted by charged particles produced by neutrino interactions in the ice.
IceCube recorded the data used for this work from 2008 to 2011 in the detector configurations \emph{IC40}, \emph{IC59} and \emph{IC79}, where the numbers denote the amounts of deployed strings.

Astrophysical neutrinos are well suitable messengers for studying production and acceleration mechanisms of cosmic rays.
Recently, IceCube discovered a diffuse all-flavor astrophysical neutrino flux \cite{bib:HESE}.
However, the properties of its astrophysical sources have not been identified yet.
Besides single point source and catalogue searches \cite{bib:ps}, an angular correlation analysis was performed on IceCube data from 2008 to 2011 to search for sources too weak to be detected individually \cite{bib:paper}.
In this work, the results of this analysis are reinterpreted with a new method to account for realistic source populations and the observed astrophysical neutrino flux.
\section{Method}
\subsection{Limits used for astrophysical interpretation\label{subsec:analysis}}
In this section, the key properties of the angular correlation analysis are summarized.
It is based on spherical harmonics expansions of muon neutrino arrival direction skymaps.
The signal expectation has been benchmarked according to different \emph{signal hypotheses}, characterized by three quantities:
The number of sources $N_{\mathrm{Sou}}$; the mean number of measured neutrinos per source at the horizon $\mu$ (\emph{source strength}); and the astrophysical spectral index $\gamma$.
Assuming a constant flux from each source, a source at the horizon is assigned a number of neutrinos following a Poisson distribution with mean $\mu$; with changing declination, the mean number of neutrinos per source is reduced according to the lower detector acceptance.
The signal part of each hypothesis includes $N_{\mathrm{Sou}}$ isotropically distributed sources in the Northern Sky with such Poisson distributed numbers of neutrinos.
To match the total number of 108310 neutrinos in agreement with the number of events in the used IceCube data sample, a background part for the respective hypothesis is added.
It consists of atmospheric neutrinos, sampled from experimental data.
A test statistic (TS) value is calculated from the spherical harmonics expansion coefficients of each skymap.
It denotes how significantly the specific skymap is distinguishable from pure atmospheric background.
By repeating this process for many skymaps of each specific signal hypothesis, a distribution of TS values is obtained for each signal hypothesis.
The shift of this distribution with respect to the distribution of the TS of the purely atmospheric background in units of the standard deviation of the distribution of the TS of the background is called \emph{signalness} $\Sigma$.
Finally, a Feldmann-Cousin approach \cite{bib:Feldmann} is used to compute experimental limits (90 \% C.L.) in terms of $N_{\mathrm{Sou}}$, $\mu$ and $\gamma$ from the TS value of the experimental skymap and the distributions of the TS of the simulated hypotheses.
\subsection{Motivation and conversion}\label{subsec:moti}
\begin{figure}
\centering
\begin{minipage}[t]{.49\textwidth}\centering
\includegraphics[width=\linewidth]{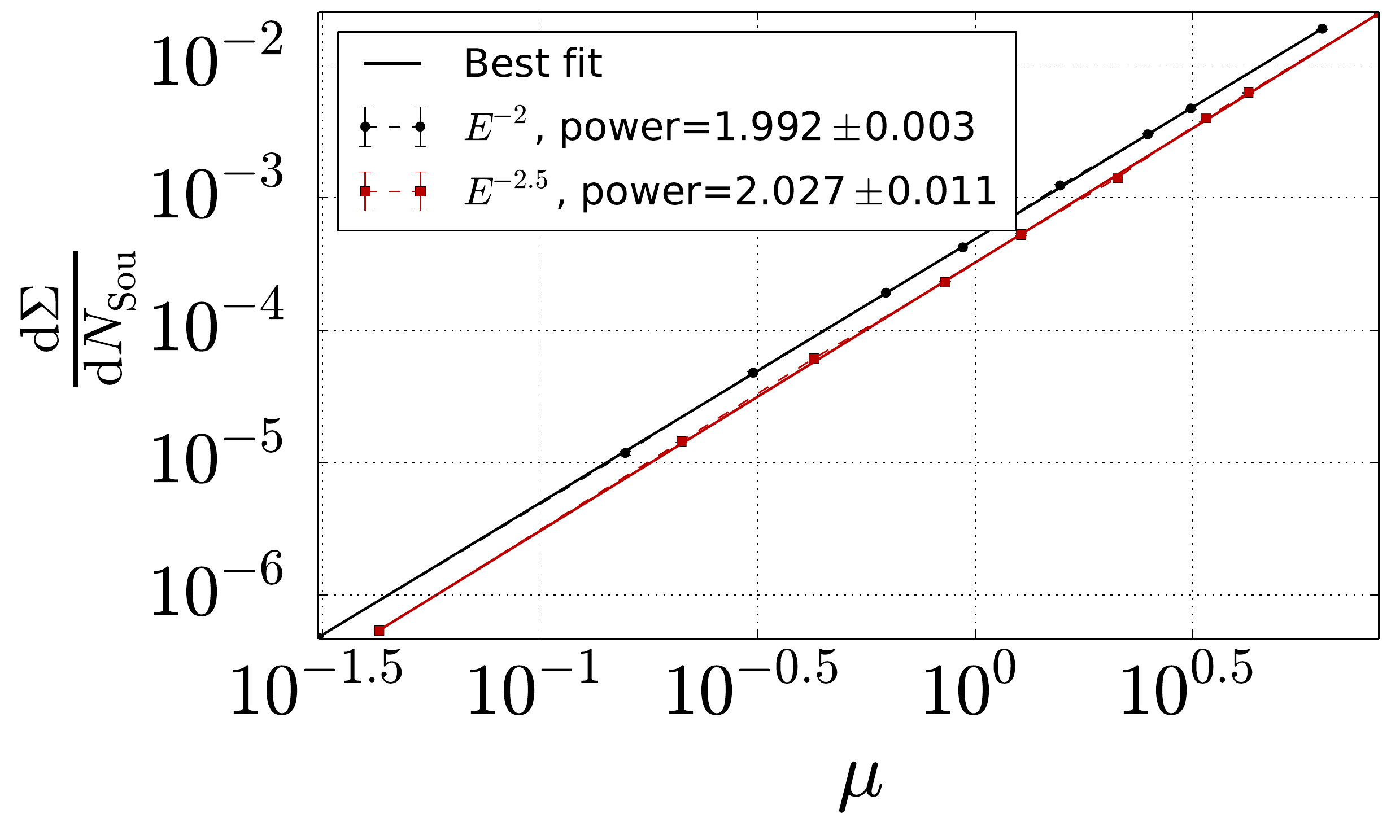}
\caption{Signalness per source $\frac{\mathrm{d}\Sigma}{\mathrm{d}N_{\mathrm{Sou}}}$ against source strength $\mu$ for astrophysical energy spectra $E^{-2}$ and $E^{-2.5}$; legend: best-fit power law exponent}
\label{fig:signalness_per_source}\end{minipage}
\hfill
\begin{minipage}[t]{.49\textwidth}\centering
\includegraphics[width=\linewidth]{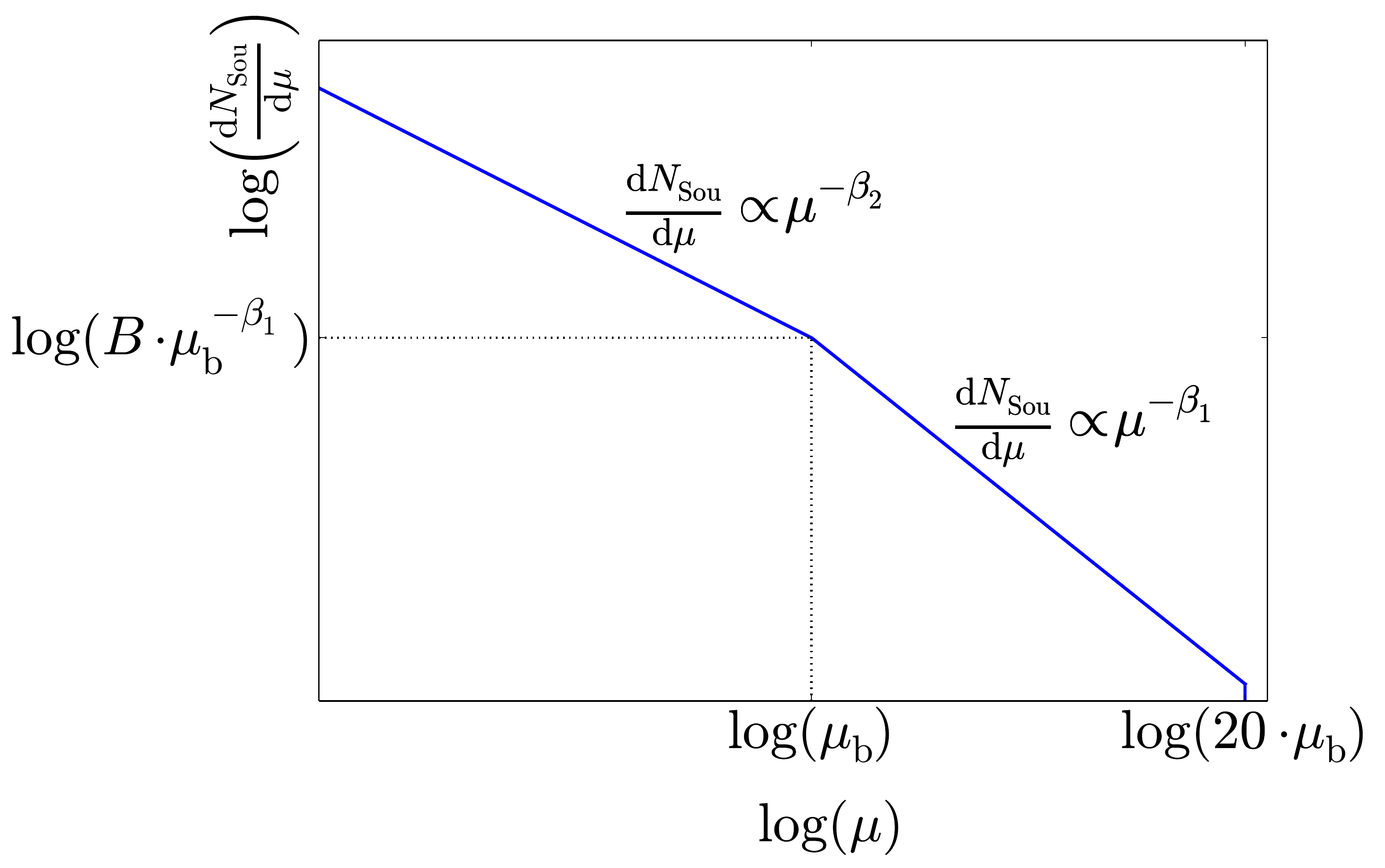}
\caption{Sketch of the source count distribution $\frac{\mathrm{d}N_{\mathrm{Sou}}}{\mathrm{d}\mu}(\mu)$ with powers $\beta_1$ and $\beta_2$ adopted from Fermi-LAT}
\label{fig:scd}\end{minipage}
\end{figure}

While the angular correlation analysis described in Section \ref{subsec:analysis} tests signal hypotheses for \emph{fixed} values of $N_{\mathrm{Sou}}$ and $\mu$, corresponding to a fixed flux per source at Earth, realistic source populations contain multiple sources of \emph{variable} strength.
Thus, to obtain limits on realistic source distributions, two possibilities exist.
First, every source population of interest can be simulated and its distribution of the TS investigated; second, the already obtained limits can be converted to limits on realistic source populations.
To describe a realistic source population, a so called \emph{source count distribution} $\frac{\mathrm{d}N_{\mathrm{Sou}}}{\mathrm{d}\mu}$ is used which describes the number of sources per (infinitesimal) source-strength interval.
The source count distribution is a function of the signal parameter $\mu$, used in the angular correlation analysis, which simplifies the conversion of the limits from the analysis to limits on realistic source distributions as shown in the following.
It should be noted that while this work focuses on the limits, also other signalness-related results from the angular correlation analysis, like the sensitivity or the discovery potential, can be converted in the same way.
To perform the conversion of the limits, two quantities have to be equated: the signalness of the limit $\Sigma_{\mathrm{lim}}$ from the angular correlation analysis ($\Sigma_{\mathrm{lim}}=1.38$ \cite{bib:paper}), and the signalness of the source population of interest, given on the right hand side of Equation \eqref{eqn:sigma}:
\begin{equation}\label{eqn:sigma}
\Sigma_{\mathrm{lim}} \overset{!}{=} \int\limits_0^{\infty}\mathrm{d}\mu\,\frac{\mathrm{d}N_{\mathrm{Sou}}}{\mathrm{d}\mu}(\mu)\frac{\mathrm{d}\Sigma}{\mathrm{d}N_{\mathrm{Sou}}}(\mu)
\end{equation}
$\frac{\mathrm{d}N_{\mathrm{Sou}}}{\mathrm{d}\mu}(\mu)$ is the source count distribution and $\frac{\mathrm{d}\Sigma}{\mathrm{d}N_{\mathrm{Sou}}}(\mu)$ is the signalness per source which is known from \cite{bib:master} to follow a power law with an exponent compatible with about 2 as shown in Figure \ref{fig:signalness_per_source}.
Thus, the signalness of the source population is the integral over the source strength $\mu$ of all sources per $\mu$ interval times their respective signalness per source.
Solving Equation \eqref{eqn:sigma} for parameters of an assumed source count distribution yields the converted limits on these parameters.
\subsection{Exemplary source count distribution parametrization adopted from Fermi-LAT\label{subsec:Fermi}}
This work focuses on extragalactic neutrino sources.
The gamma-ray telescope Fermi-LAT has detected extragalactic high-energy photon sources at high galactic latitudes \cite{bib:Fermi} with a fitted energy spectrum of $E^{-2.4}$ which are also candidates for high-energy neutrino sources \cite{bib:Elisa}. 
Thus, for the neutrino source count distribution in this work, the parametrization of the gamma ray source count distribution from Fermi-LAT is adopted.
It should be noted that this parametrization is, despite its physical motivation, only an exemplary parametrization to demonstrate the performance of this method.

The Fermi-LAT source count distribution from \cite{bib:Fermi} is the photon flux dependent number density $\frac{\mathrm{d}N_{\mathrm{Sou}}}{\mathrm{d}S}(S)$ where $N_{\mathrm{Sou}}$ is the number of sources and $S$ is the photon flux per source at Earth.
It is parametrized by a broken power law:
\begin{equation}\label{eqn:scd_original}
\frac{\mathrm{d}N_{\mathrm{Sou}}}{\mathrm{d}S} = \left\{\begin{array}{r r}AS^{-\beta_1}, & S\geq S_{\mathrm{b}} \\ AS_{\mathrm{b}}^{-\beta_1+\beta_2}S^{-\beta_2}, & S<S_{\mathrm{b}}\end{array}\right.
\end{equation}
$\beta_1 = 2.49\pm0.12$ and $\beta_2 = 1.58\pm0.08$ are the powers of the source count distribution after and before the break, respectively.
$A = \unit[(16.46\pm0.80)10^{-14}]{cm^2 deg^{-2}s}$ is a simple scale factor for $\frac{\mathrm{d}N_{\mathrm{Sou}}}{\mathrm{d}S}$ and $S_{\mathrm{b}} = \unit[(6.60\pm0.91)10^{-8}]{cm^{-2}s^{-1}}$ is the particle flux at the break of the source count distribution.
For the purpose of this work, we express this parametrization in terms of the (neutrino) source strength as defined in Section \ref{subsec:analysis} (the source strength at the break is denoted $\mu_{\mathrm{b}}$) instead of the photon flux $S$ (and a breaking flux $S_{\mathrm{b}}$) and in terms of a dimensionless scale factor $B$ instead of $A$.
All parameters we use are dimensionless in contrast to the replaced parameters.
Furthermore, a cutoff is introduced by setting the source count distribution to zero for all source strengths above a maximum $\mu_{\mathrm{max}}$ because otherwise contributions from sources with diverging strength would be included in the signalness calculation (Equation \eqref{eqn:sigma}).
It is fixed to $\mu_{\mathrm{max}} = 20\mu_{\mathrm{b}}$ corresponding to the brightest source in the Fermi-LAT sample \cite{bib:Fermi} which has a flux of $S_{\mathrm{max}}\approx 20 S_{\mathrm{b}}$.
The best-fit values for the powers of the Fermi-LAT source count distribution, $\beta_1 = 2.49$ and $\beta_2 = 1.58$, are adopted for neutrinos (except for the generalization in Section \ref{subsec:varybeta}).
This results in:
\begin{equation}\label{eqn:scd}
\frac{\mathrm{d}N_{\mathrm{Sou}}}{\mathrm{d}\mu} = \left\{\begin{array}{r r} 0, & \mu\geq20\mu_{\mathrm{b}}\\ \hfill B\mu^{-\beta_1}, & 20\mu_{\mathrm{b}}>\mu\geq\mu_{\mathrm{b}} \\ B\mu_{\mathrm{b}}^{-\beta_1+\beta_2}\mu^{-\beta_2}, &  \mu<\mu_{\mathrm{b}}\end{array}\right.
\end{equation} 
Figure \ref{fig:scd} shows an illustration of the used parametrization.
\section{Application and results}
The interpretation of the IceCube results proceeds according to the following steps:
In Section \ref{subsec:conversion}, Equation \eqref{eqn:sigma} is solved for the parameters $B$ and $\mu_{\mathrm{b}}$ of the source count distribution described in Section \ref{subsec:Fermi} using the limit signalness from the angular correlation analysis.
In Section \ref{subsec:HESE}, the parameters of the source count distribution are constrained to reproduce the expected numbers of neutrinos from the observed astrophysical neutrino flux.
In Section \ref{subsec:Fermi_epsilon}, the Fermi-LAT observation is expressed in the parameters $B$ and $\mu_{\mathrm{b}}$ as a function of a universal neutrino-to-photon ratio $\varepsilon_{\nu/\gamma}$ from all sources measured by Fermi-LAT.
Values for $\varepsilon_{\nu/\gamma}$ corresponding to the measurements of Sections \ref{subsec:conversion} and \ref{subsec:HESE} are determined.
As a last step, in Section \ref{subsec:varybeta}, the dependencies of the results for $\varepsilon_{\nu/\gamma}$ on $\beta_1$ and $\beta_2$ are investigated.
For the applications in Sections \ref{subsec:conversion}, \ref{subsec:HESE} and \ref{subsec:Fermi_epsilon}, the powers of the source count distribution, $\beta_1$ and $\beta_2$ are fixed to the Fermi-LAT best-fit values from Section \ref{subsec:Fermi}.
The neutrino energy spectral indices $\gamma = 2.0$ and $\gamma = 2.5$ are tested, motivated by IceCube measurements of the astrophysical neutrino flux \cite{bib:HESE}.
\subsection{Limit conversion}\label{subsec:conversion}
\begin{figure}
\centering
\includegraphics[width=\textwidth]{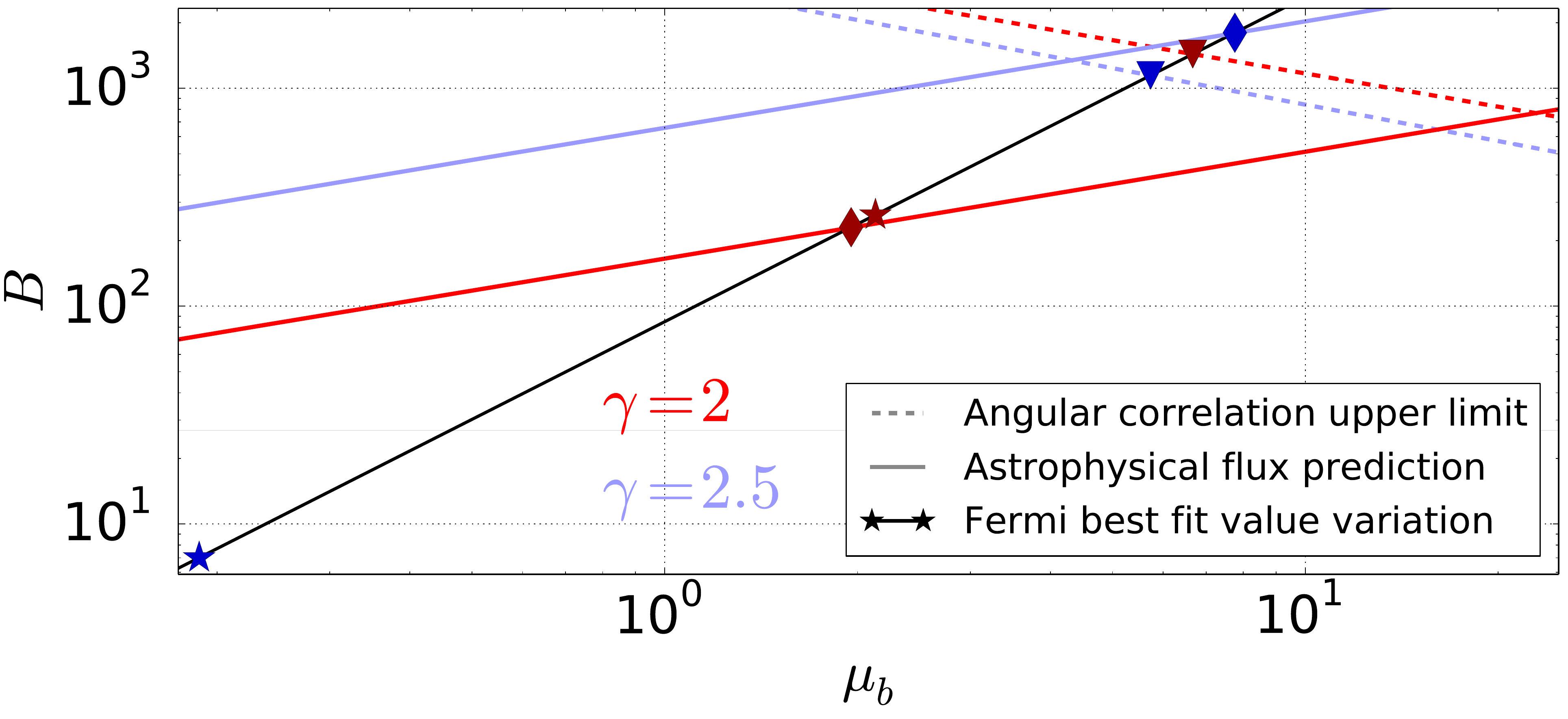}
\caption{Dashed lines: IceCube limits converted to $B(\mu_{\mathrm{b}})$; colored solid lines: $(B,\mu_{\mathrm{b}})$ reproducing the observed astrophysical neutrino flux (s. Section \protect\ref{subsec:HESE}); triangles: $(B,\mu_{\mathrm{b}})$ at limit for universal $\varepsilon_{\nu/\gamma}$; diamonds: prediction for $\varepsilon_{\nu/\gamma}$ reproducing the observed astrophysical neutrino flux purely originating from high-latitude sources measured by Fermi-LAT; asterisks: $(B,\mu_{\mathrm{b}})$ for $\varepsilon_{\nu/\gamma} = 1$; black line: allowed $(B,\mu_{\mathrm{b}})$ for universal $\varepsilon_{\nu/\gamma}$}
\label{fig:results}
\end{figure}

The solution of Equation \eqref{eqn:sigma} with $\frac{\mathrm{d}N_{\mathrm{Sou}}}{\mathrm{d}\mu}$ from Equation \eqref{eqn:scd} is a function $B(\mu_{\mathrm{b}})$.
It represents the upper limit from the angular correlation analysis \cite{bib:paper} on $B$ for each value of $\mu_{\mathrm{b}}$ and is shown as a dashed line for each spectral index in Figure \ref{fig:results}.
The negative slopes of the exclusions lines are explained by the increased signalness for larger $\mu_{\mathrm{max}}\propto\mu_{\mathrm{b}}$ (s. Equation \eqref{eqn:sigma}) due to a corresponding larger non-zero integration range.
This increased signalness is compensated by lower values for $B$, causing the negative slope.
\subsection{Constraints from the observed astrophysical neutrino flux}\label{subsec:HESE}
The astrophysical neutrino flux, observed by IceCube in 2014 \cite{bib:HESE}, can also be represented in terms of $B$ and $\mu_{\mathrm{b}}$.
The number of measured signal neutrinos expected from the source count distribution $n(\mathrm{scd})$ and the number expected from the observed flux $n(\mathrm{astroph.})$ are equated:
\begin{equation}\label{eqn:n}
n(\mathrm{scd})=\int\limits_0^\infty\mathrm{d}\mu\,\frac{\mathrm{d}N_{\mathrm{Sou}}(B,\mu_{\mathrm{b}})}{\mathrm{d}\mu}\cdot f\mu\overset{!}{=}n(\mathrm{astroph.})=\sum\limits_{\mathrm{IC}}T^{\mathrm{IC}}\int\limits_0^\infty\mathrm{d}E\, A_{\mathrm{eff}}^{\mathrm{IC}}\cdot\frac{\mathrm{d}\Phi}{\mathrm{d}E}
\end{equation}
where $\mathrm{IC}$ denotes the IceCube detector configuration, $T^{\mathrm{IC}}$ the livetime of IceCube for the operation of each detector configuration, $A_{\mathrm{eff}}^{\mathrm{IC}}$ is the effective area of each configuration and $\frac{\mathrm{d}\Phi}{\mathrm{d}E}$ the observed differential astrophysical neutrino flux by IceCube.
$f$ is a factor $\sim 1$ to obtain the declination-averaged number of measured neutrinos for a value of $\mu$.
The pairs $(B,\mu_{\mathrm{b}})$ solving this equation for the different considered neutrino energy spectra are the values for $B$ and $\mu_{\mathrm{b}}$ that correspond to the number of neutrinos expected from the observed astrophysical neutrino flux and are shown as colored lines in Figure \ref{fig:results}.
They are different since $n(\mathrm{astroph.})$ is energy spectrum dependent due to the correlation of the flux normalization and the energy spectrum.
Their intersections with the dashed limit lines (of the respective energy spectrum) separate the allowed regions (below) from the excluded regions (above) of parameter values constituting the observed flux.
\subsection{The Fermi-LAT best-fit value and the neutrino-to-photon ratio}\label{subsec:Fermi_epsilon}
Assuming a universal neutrino-to-photon ratio at Earth $\varepsilon_{\nu/\gamma}$ from all sources measured in the Fermi-LAT high-latitude survey, one can express the gamma-ray source count distribution measured by Fermi-LAT in terms of $B$, $\mu_{\mathrm{b}}$ (i.e. properties of a neutrino flux) and $\varepsilon_{\nu/\gamma}$. %Besides the examination of the results in terms of $B(\mu_{\mathrm{b}})$, the source count distribution can be parametrized using the Fermi-LAT measurement for $A$ and $S_{\mathrm{b}}$ \cite{bib:Fermi} under the assumption of a universal neutrino-to-photon ratio at Earth $\varepsilon_{\nu/\gamma}$ from the sources measured in the Fermi-LAT high-latitude survey.
First, the values for $B$ and $\mu_{\mathrm{b}}$ corresponding to an equal flux of neutrinos and photons at Earth ($\varepsilon_{\nu/\gamma} = 1$) in the energy range of the used IceCube measurement ($\unit[100]{GeV} - \unit[100]{TeV}$), have to be determined.
These particular values for $B$ and $\mu_{\mathrm{b}}$ are called $B_{\mathrm{Fermi}}$ and $\mu_{\mathrm{b,Fermi}}$.
Fermi-LAT measured the extragalactic photon flux in the energy range from $\unit[100]{MeV}$ to $\unit[100]{GeV}$.
Thus, this particle flux is extrapolated into the energy range of IceCube using the assumed energy spectra with indices $\gamma=2.0$ and $\gamma=2.5$, yielding the values for $B_{\mathrm{Fermi}}$ and $\mu_{\mathrm{b,Fermi}}$.
The extrapolation from the energy range of Fermi-LAT into the energy range of IceCube yields different flux normalizations for the energy spectra and therefore different values for $B_{\mathrm{Fermi}}$ and $\mu_{\mathrm{b,Fermi}}$.
For each value of $\varepsilon_{\nu/\gamma}$, the source strength $\mu$ of each source in the population with $B = B_{\mathrm{Fermi}}$ and $\mu_{\mathrm{b}} = \mu_{\mathrm{b,Fermi}}$ has to be multiplied by $\varepsilon_{\nu/\gamma}$.
For the source count distribution parameters $B$ and $\mu_{\mathrm{b}}$, this leads to:
\begin{equation}\label{eqn:enugamma}
\hspace{-.7cm}B=\varepsilon_{\nu/\gamma}^{\beta_1-1}\cdot B_{\mathrm{Fermi}}\hspace{3cm}\mu_{\mathrm{b}}=\varepsilon_{\nu/\gamma}\cdot\mu_{\mathrm{b,Fermi}}\hspace{-.7cm}
\end{equation}
Thus, for each value of $\varepsilon_{\nu/\gamma}$, a certain pair $(B,\mu_{\mathrm{b}})$ is obtained, resulting in the black line shown in Figure \ref{fig:results}.
The intersections of the (dashed) limit lines from Section \ref{subsec:conversion} and the (solid) lines representing the observed astrophysical neutrino flux with the (black) line representing the neutrino-to-photon ratios $\varepsilon_{\nu/\gamma}$ directly yield upper limits on $\varepsilon_{\nu/\gamma}$ and predictions for $\varepsilon_{\nu/\gamma}$ for each neutrino energy spectrum in this simplified model.
The values for the $\varepsilon_{\nu/\gamma}$ limit and the $\varepsilon_{\nu/\gamma}$ prediction are given in Table \ref{tab:results}.
\begin{table}\centering
\begin{tabular}{c c c c}
    $\gamma$ & $\varepsilon_{\nu/\gamma}$ prediction & $\varepsilon_{\nu/\gamma}$ limit & ratio \\ \hline
    2.0 & 0.92 & 3.13 & 0.29 \\
    2.5 & 41.4 & 30.6 & 1.35 \\
\end{tabular}
\caption{Results for universal $\varepsilon_{\nu/\gamma}$; 2nd column: $\varepsilon_{\nu/\gamma}$ predictions assuming the observed neutrino flux; 3rd column: $\varepsilon_{\nu/\gamma}$ (upper) limits; 4th column: ratio between $\varepsilon_{\nu/\gamma}$ prediction and $\varepsilon_{\nu/\gamma}$ limit}
\label{tab:results}
\end{table}
\subsection{Changing parameter space - varying source count distribution powers}\label{subsec:varybeta}
\begin{figure}
\centering
\begin{subfigure}{.5\textwidth}
  \centering
  \includegraphics[width=\linewidth]{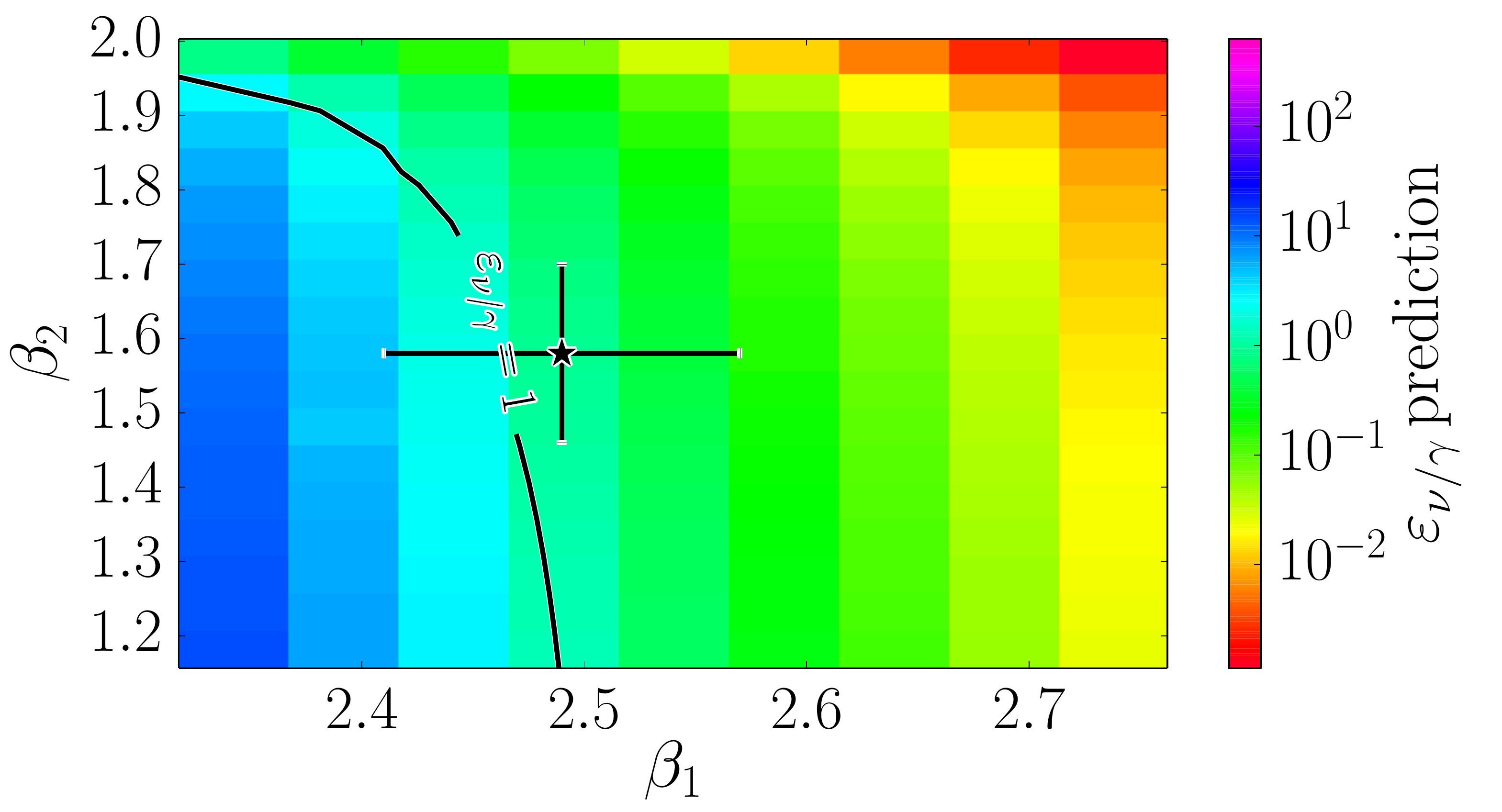}
  \caption{$\gamma=2.0$}
  \label{fig:gamma_2_pred}
\end{subfigure}%
\begin{subfigure}{.5\textwidth}
  \centering
  \includegraphics[width=\linewidth]{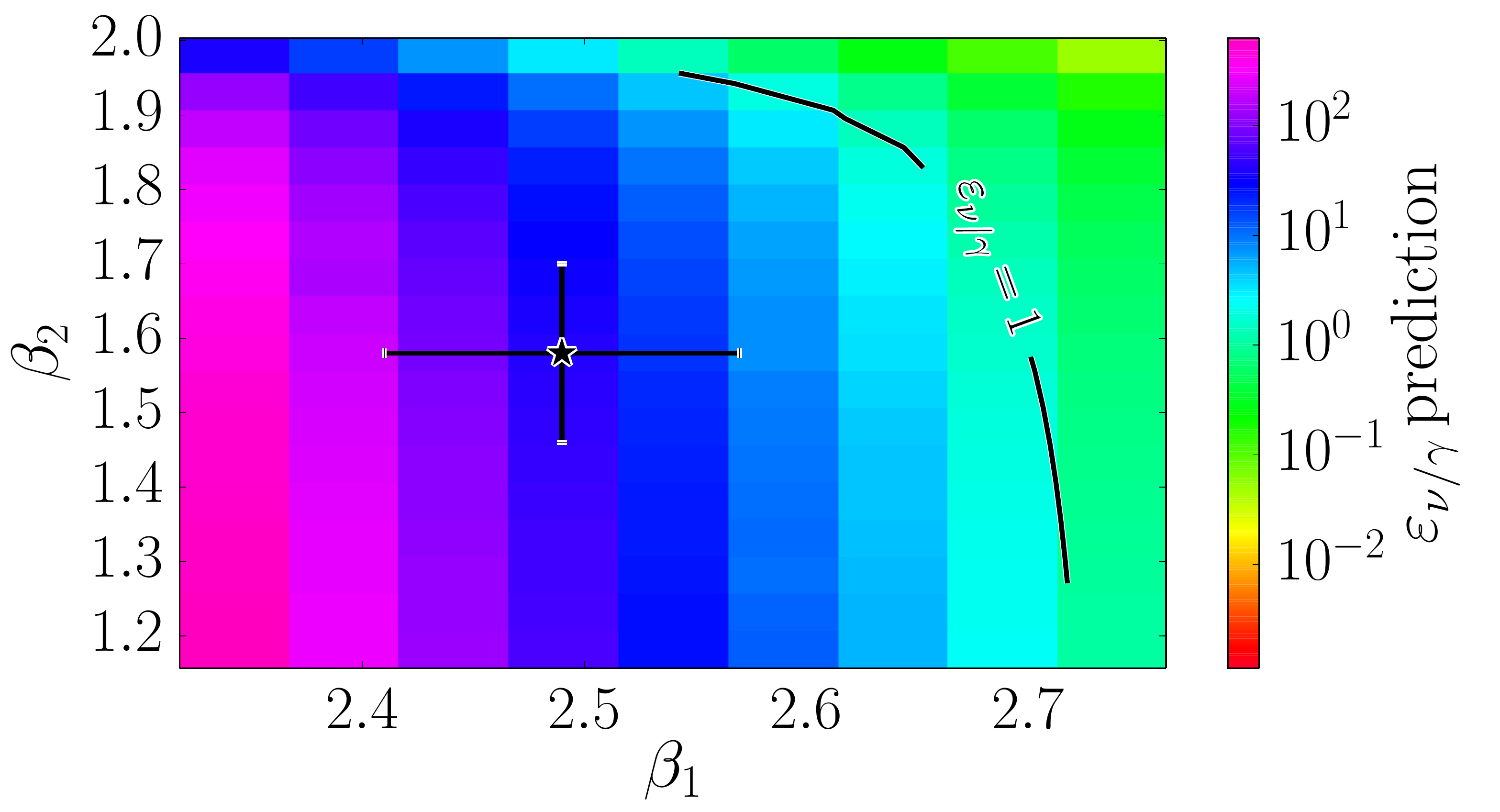}
  \caption{$\gamma=2.5$}
  \label{fig:gamma_25_pred}
\end{subfigure}
\begin{subfigure}{.5\textwidth}
  \centering
  \includegraphics[width=\linewidth]{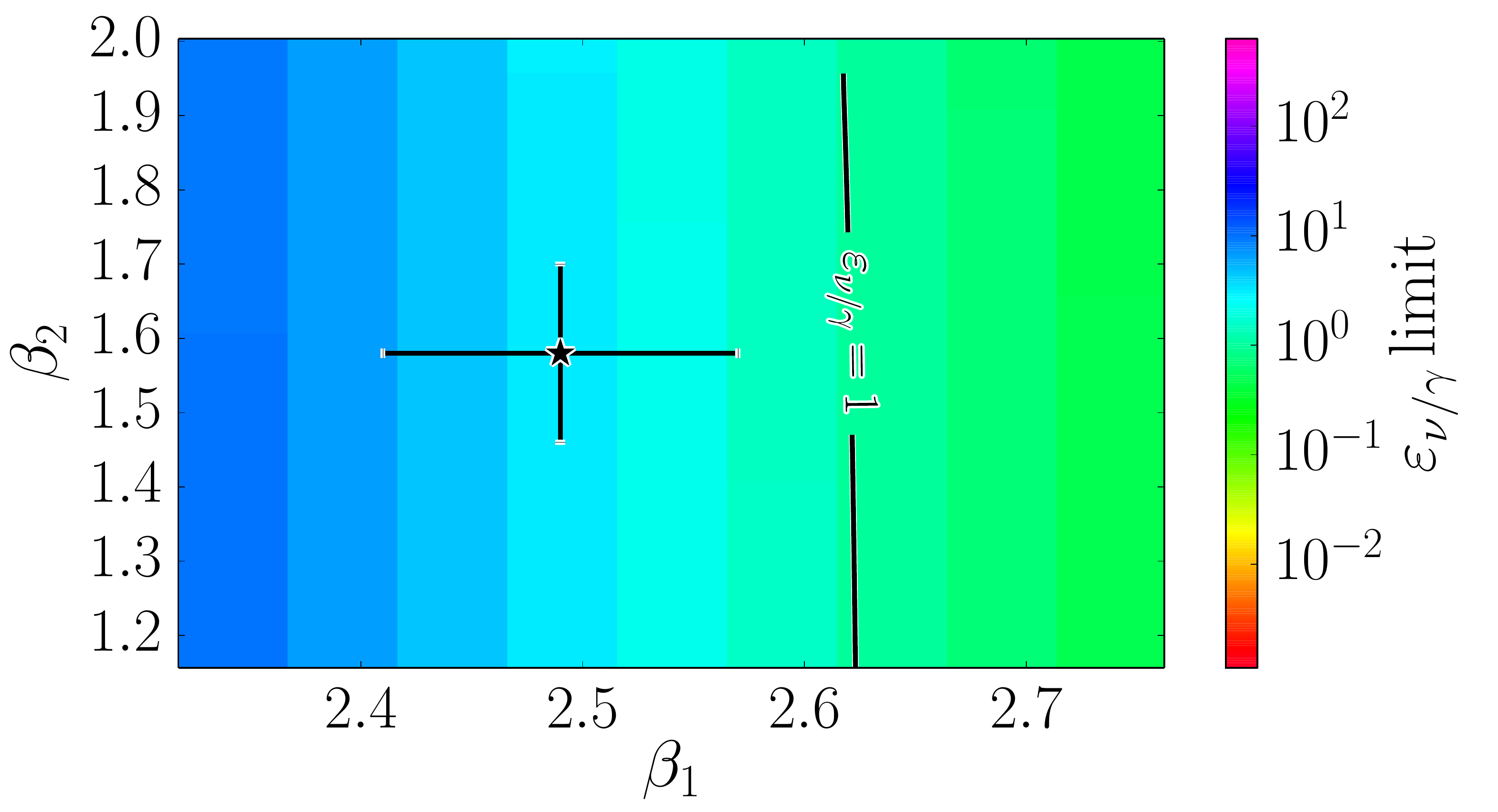}
  \caption{$\gamma=2.0$}
  \label{fig:gamma_2_lim}
\end{subfigure}%
\begin{subfigure}{.5\textwidth}
  \centering
  \includegraphics[width=\linewidth]{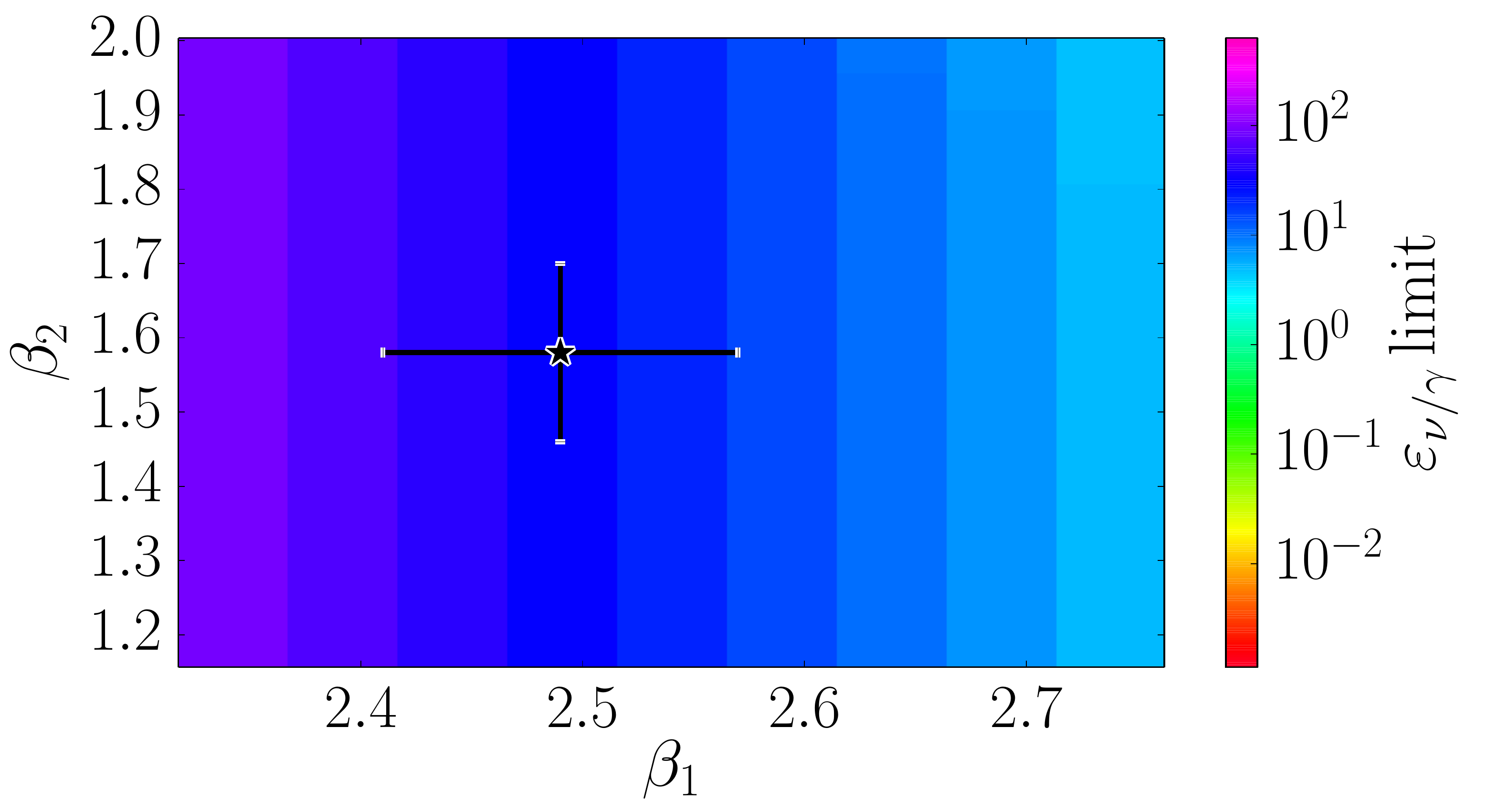}
  \caption{$\gamma=2.5$}
  \label{fig:gamma_25_lim}
\end{subfigure}
\begin{subfigure}{.5\textwidth}
  \centering
  \includegraphics[width=\linewidth]{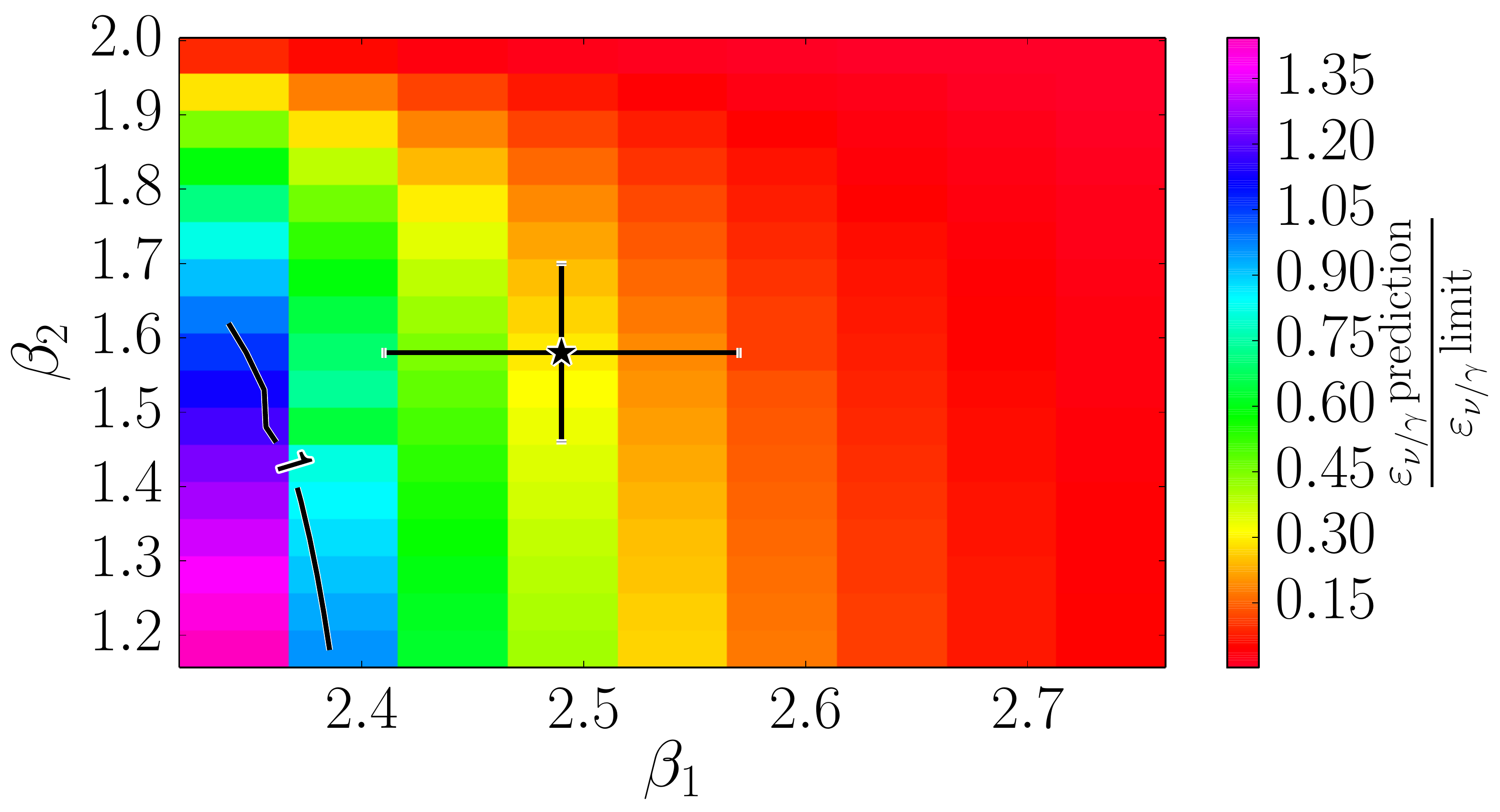}
  \caption{$\gamma=2.0$}
  \label{fig:gamma_2_ratio}
\end{subfigure}%
\begin{subfigure}{.5\textwidth}
  \centering
  \includegraphics[width=\linewidth]{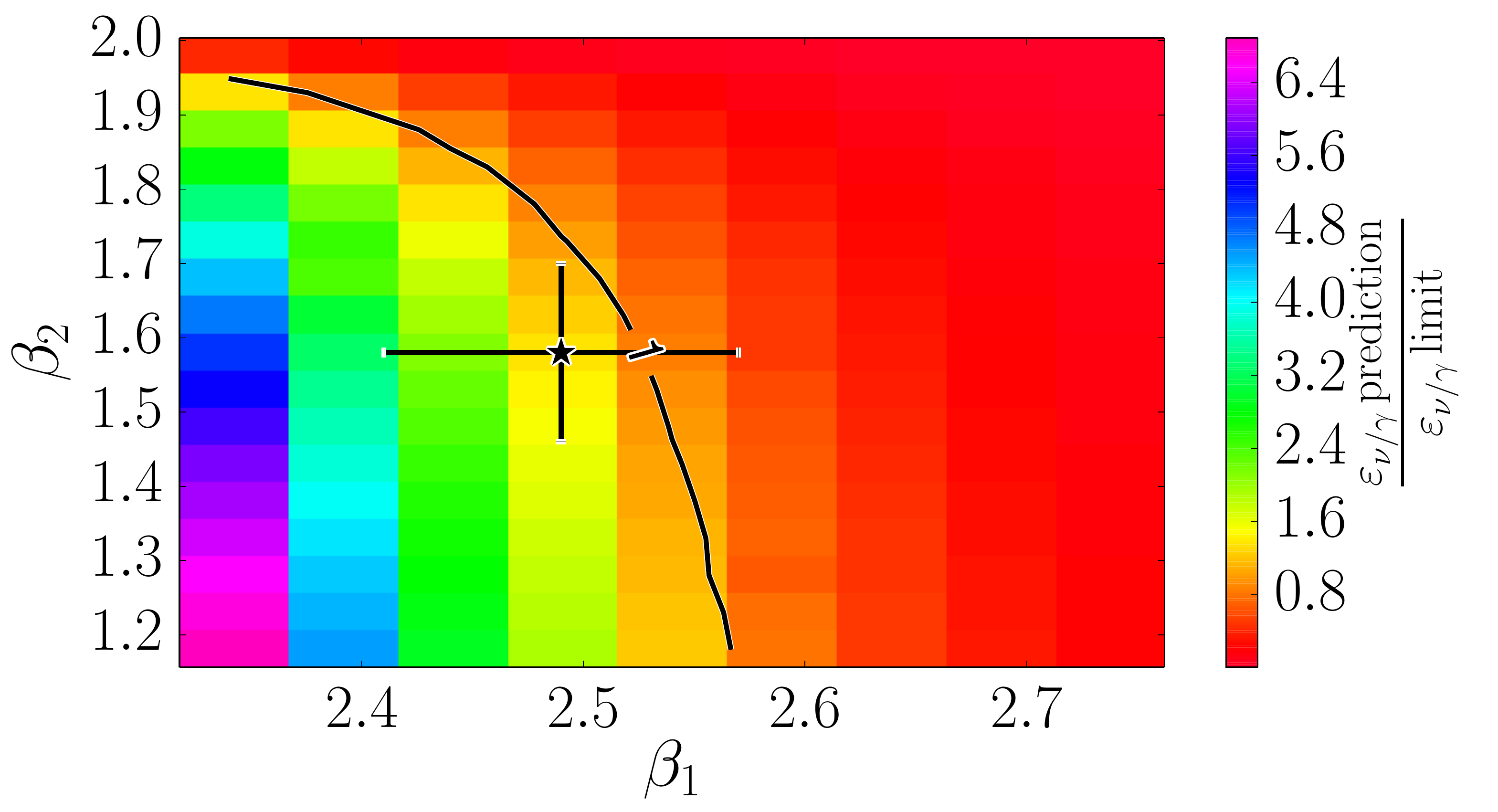}
  \caption{$\gamma=2.5$}
  \label{fig:gamma_25_ratio}
\end{subfigure}
\caption{(a), (b): predicted universal neutrino-to-photon ratios $\varepsilon_{\nu/\gamma}$ for source populations that correspond to the observed astrophysical neutrino flux for different powers $\beta_1$, $\beta_2$ (s. Equation \protect\eqref{eqn:scd}); (c), (d): converted limits on universal $\varepsilon_{\nu/\gamma}$ for different powers $\beta_1$, $\beta_2$ (s. Equation \protect\eqref{eqn:scd}); (e), (f): ratios between the $\varepsilon_{\nu/\gamma}$ prediction and the $\varepsilon_{\nu/\gamma}$ limit; asterisk with error bars: $\beta_1$, $\beta_2$ and uncertainties from Fermi-LAT \cite{bib:Fermi}; black line: color-coded quantity $= 1$}
\label{fig:scan}
\end{figure}

The procedure from Section \ref{subsec:conversion} through \ref{subsec:Fermi_epsilon} is repeated with varied powers of the source count distribution $\beta_1$ and $\beta_2$ (defined as in Equation \eqref{eqn:scd}).
Figures \ref{fig:gamma_2_pred} and \ref{fig:gamma_25_pred} show the results in terms of the $\varepsilon_{\nu/\gamma}$ prediction as introduced in Section \ref{subsec:HESE} and Figures \ref{fig:gamma_2_lim} and \ref{fig:gamma_25_lim} show the $\varepsilon_{\nu/\gamma}$ limit as introduced in Section \ref{subsec:Fermi_epsilon} for the neutrino energy spectra with $\gamma=2.0$ and $\gamma=2.5$.
Figures \ref{fig:gamma_2_ratio} and \ref{fig:gamma_25_ratio} show the ratios between the $\varepsilon_{\nu/\gamma}$ prediction and the $\varepsilon_{\nu/\gamma}$ limit.
Ratios larger than $1$ indicate that the astrophysical neutrino flux is excluded with 90 \% C.L. to origin purely from source populations with the corresponding source count distributions with universal $\varepsilon_{\nu/\gamma}$ in the respective energy ranges denoted in Section \ref{subsec:Fermi_epsilon}.
\section{Conclusions}
The universal neutrino-to-photon ratios $\varepsilon_{\nu/\gamma}$ resulting in a number of neutrinos corresponding to the observed astrophysical neutrino flux by IceCube, called the $\varepsilon_{\nu/\gamma}$ prediction, and the limit on $\varepsilon_{\nu/\gamma}$, set by the angular correlation analysis in \cite{bib:paper} are summarized in Table \ref{tab:results}.
For $\gamma=2.0$, the observed astrophysical neutrino flux corresponds to a lower $\varepsilon_{\nu/\gamma}$ prediction than the $\varepsilon_{\nu/\gamma}$ limit.
Thus, for this energy spectrum, the sources from the Fermi-LAT high-latitude survey are not excluded to be the origin of the astrophysical neutrino flux under the stated assumptions.
For $\gamma=2.5$, the opposite is the case: The $\varepsilon_{\nu/\gamma}$ prediction exceeds the $\varepsilon_{\nu/\gamma}$ limit and is thus excluded.

The results of the variations of $\beta_1$ and $\beta_2$ (s. Equations \eqref{eqn:scd_original} and \eqref{eqn:scd}) reveal several insights:
First, $\beta_1$ plays a major role for both, the $\varepsilon_{\nu/\gamma}$ prediction and the $\varepsilon_{\nu/\gamma}$ limit since the source count distribution depends strongly on $\beta_1$ for all source strengths $\mu$.
Second, the $\varepsilon_{\nu/\gamma}$ limit is almost independent of $\beta_2$ as shown in Figures \ref{fig:gamma_2_lim} and \ref{fig:gamma_25_lim}.
As $\beta_2$ only affects the source count distribution $\frac{\mathrm{d}N_{\mathrm{Sou}}}{\mathrm{d}\mu}$ for source strengths below the source strength of the break $\mu_{\mathrm{b}}$ (s. Equation \eqref{eqn:scd}), this leads to the conclusion that the sources brighter than $\mu_{\mathrm{b}}$ are the signalness dominating sources.
In contrast to that, for the $\varepsilon_{\nu/\gamma}$ prediction corresponding to the astrophysical neutrino flux, there is a remarkable dependence on $\beta_2$.
The quantity used for the $\varepsilon_{\nu/\gamma}$ prediction is the number of neutrinos from the tested source count distribution $n(\mathrm{scd})$.
As the integrand in the formula for $n(\mathrm{scd})$ (Equation \eqref{eqn:n}) is one power of $\mu$ lower compared to the integrand in the formula for the signalness (Equation \eqref{eqn:sigma}), fainter sources (i.e. small $\mu$) affect the $\varepsilon_{\nu/\gamma}$ prediction while barely affecting the $\varepsilon_{\nu/\gamma}$ limit.
A ratio between the $\varepsilon_{\nu/\gamma}$ prediction and the $\varepsilon_{\nu/\gamma}$ limit larger than 1 is excluded with 90 \% C.L.
Thus, for both examined energy ranges, the areas to the lower left from the black lines in Figures \ref{fig:gamma_2_ratio} and \ref{fig:gamma_25_ratio} are excluded.
For $\gamma=2.5$, where the hypothesis with the benchmark values for $\beta_1$ and $\beta_2$ is excluded, their uncertainty intervals reach into the allowed region.
\acknowledgments
This work is supported by the Federal Ministry of Education and Research (BMBF), the Helmholtz Alliance for Astroparticle Physics (HAP) and the German Research Foundation (DFG).
We thank Markus Ahlers, Jan Auffenberg and Leif R\"adel for the valuable discussions during the preparation of this manuscript.

\end{document}